\documentclass[12pt, letterpaper, fleqn]{article}

\usepackage[magyar]{babel}
\usepackage{titlesec}
\usepackage{amsmath}
\usepackage{amsfonts}
\usepackage{fancyhdr}
\usepackage{enumitem}
\usepackage{xpatch}
\usepackage{amssymb}
\usepackage{amssymb}
\usepackage{amsmath}
\usepackage{amscd}
\usepackage{enumerate}
\usepackage{array}
\usepackage{tabularx}
\usepackage{t1enc}
\usepackage[mathscr]{eucal}
\usepackage{graphicx}
\usepackage[utf8]{inputenc}

\begin{document}

\centerline{\large Statistical complexity of the kicked top model considering chaos}

\vspace{0.5cm}

\centerline{A. F\"ul\"op}

\vspace{0.3cm}

\centerline{University E\"otv\"os Lor\'and}

\vspace{1.0cm}

{\large \bf Abstract}

\vspace{0.3cm}

The concept of the statistical complexity is studied to characterize the classical kicked top model which plays important role in the qbit systems and the chaotic properties of the entanglement. This allow us to understand this driven dynamical system by the probability distribution in phase space to make distinguish among the regular, random and structural complexity on finite simulation. We present the dependence of the kicked top and kicked rotor model through the strength excitation in the framework of statistical complexity.  


\section{\bf Introduction}

In this article we study the driven systems considering the statistical complexity.
The concept of statistical complexity has been introduced in different way from complexity of finite series (Lempel, Ziv)\cite{lz}, algorithmic complexity (Kolmogorov)\cite{ank}, the amount of information about the optimal prediction, where the future fulfills to the expected past (Crutchfield, Young) \cite{cy}

The effective entropy was published by Grassberger \cite{gpijt} considering the mixture of the order and disorder, regularity and randomness, since the entropy of most systems is between the maximum and minimum entropy values. 

The definition of statistical complexity was introduced by L\'opez, Ruiz, Manchini, Calbet (LMC)\cite{ap} and Shiner, Davison, Landsberg (SDL) \cite{plx}.
The generalized statistical complexity measure (Martin, Palestino, Rosso)\cite{mpr} is based on the LMC's that gives a description of the finite sequence of nonlinear systems with the adequate probability distribution of the time dependent method. It was extended to Tsallis, Wootters, Renyi entropy and Kullback-Leibler,  Jensen-Shannon divergence. Tsallis suggested a generalization of the  Shannon-Boltzmann-Gibbs entropy measure \cite{ct}.  The new entropy funtional plays an significant role along with its corresponding thermodynamics (1998).  Wootters reflected on  the Euclidean distance \cite{wkw}, because he studied this concept in a quantum mechanical field; this  consideration allowed to consider an intrinsic statistical measure, this concept can be employed to any probability space.

Both experimental and theoretical sign can be evaluated by  the information theory tools, as entropy, distance, statistical divergence provides an opportunity to make estimation, detection and transmission processes.  

In the last two decades more kind of complexity measures and methodologies were introduced with their time evolution connected to optimal predictability, symbolic analysis\cite{ac}, algorithmic data compression, number system, pseudorandom bit generator, earthquake  the chaotic regim etc. \cite{lz,cy,wwaks,lmc,fc1,plx,mpr,lmpr,av,afs,cghl,lo,fpp}.
The dynamics of the statistical complexity measure is formulated according to the second law of thermodynamics. According to, entropy increases monotonically at time. It follows that the quantity $H$ is implied as an arrow of time.

We study the kicked top and kicked rotor model in this article, these driven systems are intensively researched field in quantum mechanics. These are studied in the field of quantum chaos and entanglement in ergodic and non ergodic system \cite{plq}.
 In the past two decades  the kicked top model was  an intensively researched area which contained the chaotic dynamics and quantum correlations considering  in quantum information and computation. The kicked top is a suitable model for studying spins or qbits and corresponds for the studying of entanglement. We approximate the classical limit of the kicked top if  the number of spins tend to infinity. 
 
Therefore this model is an important area of research \cite{bs2,ms4,mdl1,rlp2} for the study of entanglement  \cite{lm2,gsjls,ms3,l3,bl3,wgsb} and its relationship to classical dynamics\cite{sg3}, sign of bifurcations on different quantum correlation measures\cite{bs2}, quantum classical transition with respect to periodic trajectories \cite{ms4} and the behavior of entropy  
in the transition to chaos cite{zs}. Measure of quantum correlations  is strongly correlated with the qualitative nature of classical phase space,  whether it is regular or chaotic \cite{bs2,rlp2,lm2,bl1,mgtg,fmt1,zs}. The importance of the kicked top model is also demonstrated by the large number in a series of papers \cite{csagj,bs2,rlp2,lm2,bl1,mgtg,fmt1,nrfck,bs}.

The structure of the article contains the next parts: 

In the section (\ref{sec-stat}) we introduce the idea of complexity accordingly the measure of entropy   and disequilibrium with the probability distributions  by the by LMC functional associating to SCM family.  We discuss the statistical complexity considering the  Wootters, Kullback-Leibler  relative entropy and the Jensen-Shannon divergency. The time evolution of the SCM associated to the evolving of entropy. In the section (\ref{sec-top}) the quantum kicked top  model is investigated by the Hamiltonian functions considering the properties of  the classical equation motions. The quantum kicked rotor system is derived by the Hamiltonian and we compere the chaotic behavior of these systems in the section (\ref{sec-rot}). The Numerical approximation of the statistical complexity of the models is discussed in the section (\ref{sec-num}).

\section{Statistical Complexity Measures}\label{sec-stat}

In this section we discuss the entropy and distance  in the probability space which can be used to determine the statistical complexity measure. This plays important role in the dynamics of quantum-classical transition and the chaotic motion. We review the main futures of statistical  considerations describing dynamical properties.

\subsection{Information Measures}

The information measure $I$ is defined by a given probability distribution.  $I[P]$ refers as the  measure of the uncertainty connected to  probability distribution $P=\{p_j,j=1\dots N\}$, where $N$ indicates the  the number of possible states of the systems satisfying $\sum_{j=1}^Np_j=1$ (micro-canonical ensemble).

If $I[P]=I_{min}=0$ then this means that the maximum information is extracted from all possible outputs states.
 Otherwise the ignorance appears when   $I[P]=I[P_e]\equiv I_{max}$; $P_e=\{p_i=1/n; \forall i\}$, $P_e$ being the uniform distribution. These are the trivial cases. We define  the amount of disorder $H$ at a given  probability distribution $P$ and considering the information measure $I[P]$:
\begin{eqnarray}
H[P]=I[P]/I_{max}
\end{eqnarray}
The value of $H$ is changing $0\le H \le 1$.

Based on the Shannon-Kinchin paradigm $I$ is introduced in the expression of entropy. It can arise by canonical formulation (Boltzmann-Gibbs) of statistical mechanics which is expandable  to another entropy term as Renyi, Tsallis \cite{rmlmp}. We define the disorder $H$ for  the $P\equiv \{p_i, i=1\dots N\}$ on a discrete probability distribution: 
\begin{eqnarray}
H[P]=S[P]/S[P_e],
\end{eqnarray}
where $S[P]$ means Shannon's logarithm entropy \cite{cs} by this form 
\begin{eqnarray}
S[P]=-\sum_{j=1}^N p_j\log(p_j)
\end{eqnarray}
and $S[P_e]=\log N$.

\subsection{Distances and Statistical Complexity Measure}

In order to define Statistical Complexity Measure (SCM)  we need  to use   some distance $D$ \cite{entr} between given $P$ and the uniform  distribution $P_e$ on the available states of the system \cite{lmc,mpr,lmpr}.
\begin{eqnarray}
 Q[P]=Q_0\cdot D[P,P_e],
\end{eqnarray}
where $Q_0$ is a normalization constant ($0\le Q\le 1$) i. e. the inverse of the maximum   distance $D[P,P_e]$. The value of largest distance corresponds to that one component  of  probability distribution $P$ takes 1 and the others equal to zero.  The disequilibrium-distance $Q$ shows the structure of the system, because the "privileged" states differ from zero probability value.  

The functional form of the SCM is introduced by Lopetoz-Ruiz, Manchini and Calbet (LMC) \cite{lmc}.
\begin{eqnarray}
C[P]=H[P]\cdot Q[P]
\end{eqnarray}
This quantity represents at a given scale between  the amount of information stored in the system and its disequilibrium \cite{lmc}. In this article we study complex dynamics where the different regime are mixed, i.e. chaos, regular islands and trajectory that are neither periodic nor chaotic can be featured by SCM. 

Different distance-forms $D$  can be used to define the quantity $Q$ for the SCM.    
In the following we apply two discrete probability distributions  $P_i\equiv \{p_1^{(i)}\dots p_N^{(i)}\}$, with $i=1,2$  consider the next options:

(I) Euclidean norm $D_E$ in $\mathbb{R}^N$ \cite{lmc}:

This is  natural case for the distance $D$. We get
\begin{eqnarray}
D_E[P_1,P_2]=\|P_1-P_2\|_E^2=\sum_{j=1}^N\left\{p_j^{(1)}-p_j^{(2)}\right\}^2
\end{eqnarray}

This is the disequlibrium  term which was contained in the original  complexity measure  by L\'opez-Ruiz, Manchini and Calbet (LMC-complexity measure \cite{lmc}). 
Wootter extended it to the possibility, where we also consider the shape of the probability distributions.

(II) Wootter's distance $D_W$ \cite{mpr,wkw}

The concept of statistical distance is extended in a quantum mechanical field. He proposed a definition to distinguish among different preparations of a given quantum state and to take it into account that two such states differ from one another inside statistical error. It allows to   consider  an intrinsic statistical nature, this can be applied to any probabilistic space \cite{wkw}.
\begin{eqnarray}
D_W[P_1,P_2]=\cos^{-1}\left\{\sum_{j=1}^N \left(p_j^{(1)}\right)^{1/2}\cdot \left(p_j^{(2)}\right)^{1/2}\right\}
\end{eqnarray}
Two divergence classes were distinguished by Basseville \cite{bas}.  The first class contains divergences defined by relative entropy, while the second one pays attention  divergences related as entropy differences. 

(III) Kullback-Leiber relative entropy $D_K$ \cite{kl1}:

The relative entropy of $P_1$ with respect to $P_2$ connected to Shannon measure is the relative Kullback-Leibler Shannon entropy in the discrete case follows
\begin{eqnarray}
D_K[P_1,P_2]=K[P_1|P_2]=\sum_{j=1}^N p_j^{(1)}\log \left(\frac{p_j^{(1)}}{p_j^{(2)}}\right)
\end{eqnarray}
The distance between the probability distribution $P$ and uniform distribution $P_e$ in the Kullback-Leibler Shannon expression is given by this form
\begin{eqnarray}
D_K[P,P_e]=K[P|P_e]=S[P_e]-S[P]
\end{eqnarray}

 (IV) Jensen divergence $D_j$ \cite{lmpr}:
  
 The entropic difference $S[P_1]-S[P_2]$ does not mean an information gain (or divergence),  bacause the difference is not inevitably positive definite. Jensen's divergence is a symmetric version of the Kullback-Leibler relative entropy, which can be  written  in the form of the Shannon entropy  as follow:
 \begin{eqnarray}
 \begin{array}{rl}
 D_J[P_1,P_2]&= J_S[P_1,P_2]=\{K[P_1|P_2]+K[P_2|P_1]\}/2\\
&=S[\frac{P_1+P_2}{2}]-S[P_1]/2-S[P_2]/2
\end{array}
\end{eqnarray}
The Jensen-Shannon divergence verifies the following properties
\begin{eqnarray}
\begin{array}{rl}
(i)& J_S[P_1,P_2]\ge 0\\
(ii)& J_S[P_1,P_2]=J_S[P_2,P_1]\\
(iii)& K_S[P_1,P_2]=0 \Leftrightarrow P_2=P_1
\end{array}
\end{eqnarray}
It square root fulfills the triangle inequality:
\begin{eqnarray}
(iv)\;\;\; (J_S[P_1,P_2])^{1/2}+(J_S[P_2,P_3])^{1/2} = (J_S[P_1,P_3])^{1/2}
\end{eqnarray}
 So  the square root of the Jensen-Shannon divergence is a metric \cite{bh1}.
These entropy concepts  are extensive quantities in thermodynamics, therefore the associated statistical complexity will be an intensive quantity.

Generally  on the basis of LMC-functional product term we get a family of SCMs for each four disequilibrium
\begin{eqnarray}
C^{\nu}[P=H[P]\cdot Q_{\nu}[P]
\end{eqnarray}
The index $\nu=E,W,K,J$ denoted the disequilibrium distance which is determined with the adequate distance measure (Euclidean, Wootters, Kullback-Leibler, and Jensen-Shannon)
Then the SCM family for $\nu=K$ is following
\begin{eqnarray}
C^{(K)}[P]=H[P]\cdot Q_K[P]=H[P]\cdot (1-H[P])
\end{eqnarray}
The generalized functional term  was introduced by  Davison and Landsberg \cite{plx} for the SCM. Similar results are published by these article \cite{cfs1,bp1,sdl1}.

We consider  three members of the family $C^\nu$ $(\nu=E,W,J)$ these are not trivial functions of the entropy \cite{mpo} because they associate to two dissimilar probabilities distributions  $P$ and  uniform distribution $P_e$.
It can be seen that a given $H$ value determines a range of SCM values from $C_{min}$ to $C_{max}$. These bounds are changing during the time evolution.  
  We obtain the range $C_{min}$ and $C_{max}$ relating to the generalized $C^{\nu}=H\cdot Q_\nu$ family   which provides more information corresponding to the correlation structure between the elements of physical system. 

\subsection{The Evolution}

In statistical mechanics isolated systems \cite{cl}  play important role  featured by an initial   discrete probability distribution going toward  equilibrium.  The uniform distribution $P_e$ characterizes the equilibrium. The evolution of  the SCM can be plotted on the Figure of $C$ versus time $t$.  Nevertheless  in isolated system the entropy grows monotonically with time $(dH/dt\ge 0)$  by  the second law of thermodynamics \cite{pp1}. It follows that $H$ behaves as an arrow of time, i.e. the time evolution of the SCM corresponds to plot $C$ versus $H$.  The normalized entropy-axis equivalent  with the time-axis. \cite{lmc,rmlmp,rlmpf}

\section{Kicked top model}\label{sec-top}

\paragraph{Quantum Kicked top}

The Quantum Kicked top (QKT) is e time-dependent periodic system, which is described  by an angular momentum vector $J=(J_x,J_y,J_z)$. 
Here we choose natural unit where the Planck's constant has been adjusted to unity.
The  time evolution of the model is given by Hamiltonian  
\begin{eqnarray}\label{ht}
H(t)=pJ_y+\frac{k}{2j}J_z^2 \sum_{n=-\infty}^{\infty}\delta(t-n\tau)
\end{eqnarray}
The first expression of the Equation (\ref{ht}) means the free precession of the kicked top model around $y$ axis with angular frequency $p$. The second expression indicates the periodic $\delta$ kicks on the kicked top system, where each kick causes a torsion by an angle  $(k/2j)J_z$ about the $z$ axis. 
The components of angular momentum satisfy  the commutation relations in standard algebra of angular momentum:
\begin{eqnarray}\label{komr}
[J_i,J_j]=i\varepsilon_{i,j,k}J_k
\end{eqnarray}
The magnitude of total angular momentum $J^2=j(j+1)\hslash^2$  is conserved quantity. The classical limit is obtained when $j\to \infty$.
The time between periodic kicks corresponds to $\tau$.
 In this article it is chosen unit ($\tau=1$). 
 The parameter $k$ characterizes the chaotic behavior of the system and the strength of the kick.  
  If $k=0$ then the equation (\ref{ht}) can be integrated, which is the classical boundary  of the system.  As the value of $k$ increases, the chaoticity of this model is growing.  
   
   The expression of the  periodic-one Floquet operator for the Hamiltonian equation (\ref{ht}) is as follows:  
\begin{eqnarray}
U=\exp\left(-i\frac{k}{2j}J_z^2\right)\exp(-ipJ_y)
\end{eqnarray}
Each time period contains a linear rotation by angle $p$ around the $y$ axis and a nonlinear rotation around the $z$ axis. 
The dimension of Hilbert space is  $2j+1$ therefore the time dependent behavior can be described without any  truncation of the Hilbert space.

The quantum simulation of a given set of qbits  with $N=2j$ is well described by the quantum  kicked top model for given angular momentum  $j$. These are half-spin particles which are confined to a subspace that is symmetrical for qbit exchange.

In the symmetric subspace the state vector is defined by the following states $\{|j,m>:(m=-j,-j+1,\dots ,j)\}$ where $j=N/2$. The ground conditions fulfill the following conditions $S_z|j,m>=m|j,m>$ and $S_{\pm}|j,m>=\sqrt{(j\mp m)(j\pm m+1)}|j,m\pm 1>$, where $S_z$ and $S_{\pm}$ are collective spin operator \cite{mc}.
This is a multiqubit system and the collective properties evolve according to Hamiltonian Eq (\ref{ht}).

According to the quantum mechanics, the initial state of the kicked top system is a spin coherent state(minimum-uncertainty states) pointing in the direction of $\Phi,\Theta$. The time evolution is determined by the Floquet operator. 
The classical map of the kicked top model is discuss below.

\begin{figure}
\begin{center}
\includegraphics[width=6.0cm]{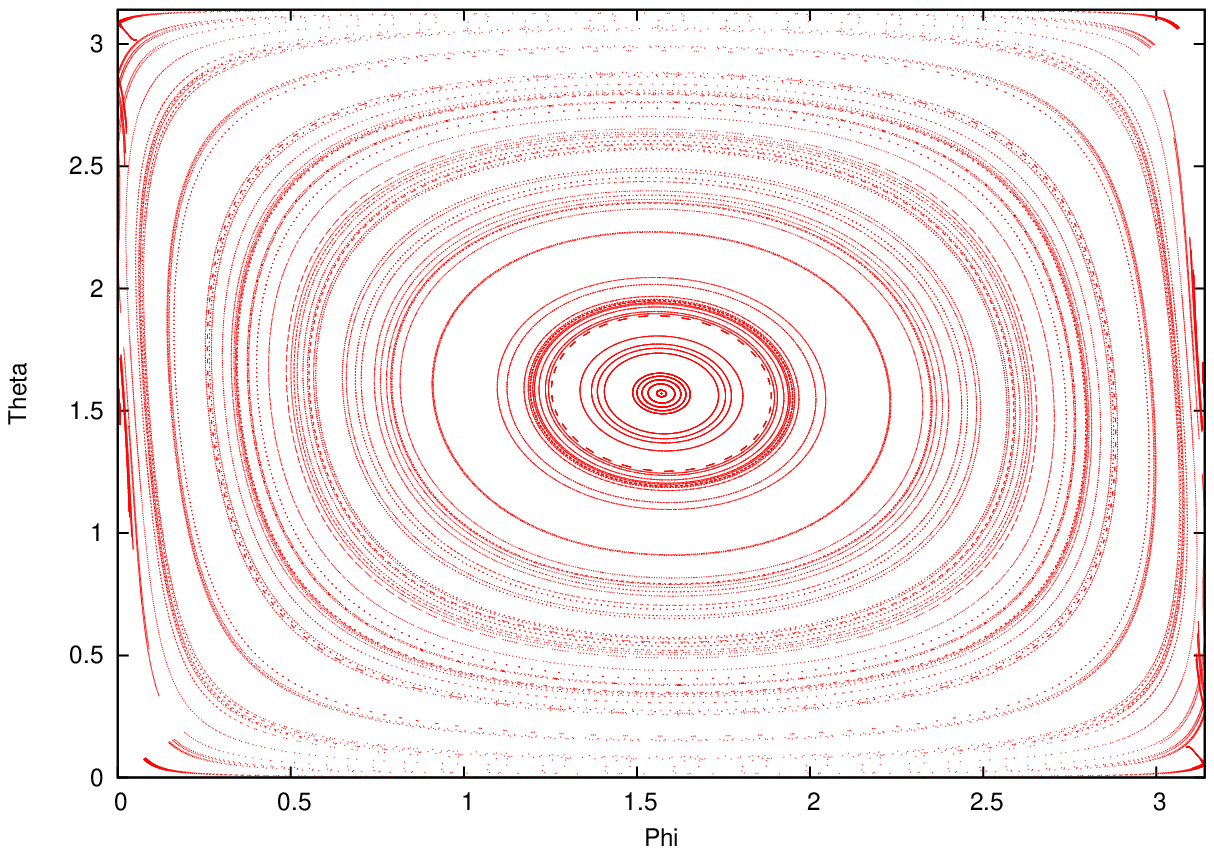}
\includegraphics[width=6.0cm]{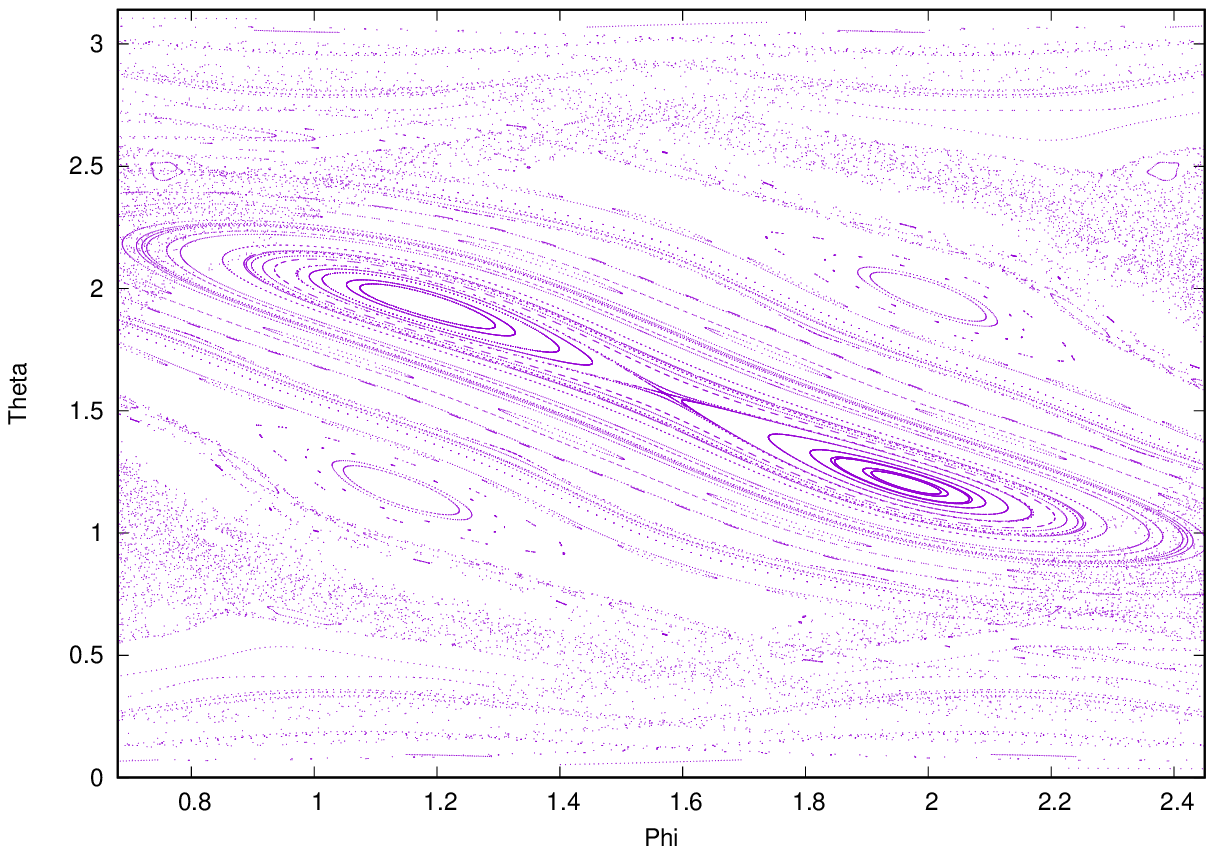}
\caption{The $\Theta$ depends on $\Phi$ variables at the kicked top model at $k=0.07$, $k=2.2$.}\label{figw-2}
\end{center}
\end{figure}

 \paragraph{Classical Kicked Top}
 
The phase space is represented in Fig. (\ref{figw-2}) as a function of coordinates $\Phi$ and $\Theta$.
 
The classic map of the kicked top model is derived in the following form \cite{hks}
\begin{eqnarray}\label{me}
\begin{array}{rl}
X'&=(X\cos p + Z\sin p)\cos [k(Z\cos p-X \sin p)]-Y\sin(k(Z\cos p- X\sin p))\\
Y'&=(X\cos p+Z\sin p)\sin[k(Z\cos p-X\sin p)]+Y\cos [k(Z\cos p-X\sin p)]\\
Z'&=-X\sin p+ Z\cos p
\end{array}
\end{eqnarray}
The  time-dependent variables $(X,Y,Z)$ fulfill the constraint $X^2+Y^2+Z^2=1$.  The trajectories are located on a sphere of unit radius.
 
 These equations (\ref{me}) can be specified by polar coordinates i.e. with polar angle $\Phi$ and   azimuth angle $\Theta$, therefore   $X=\sin\Theta \cos \Phi, Y=\sin\Theta \sin \Phi, Z=\cos \Theta$. During the time evolution of the equations, the values of $\Phi$ and $\Theta$ are determined in each step.
The symmetry properties of the model are discussed below.

The phase space is reflective on  $\Theta=\frac{\pi}{2}$ during the transformation $k\to -k$.
This is fulfilled because $k\to -k$ transformation has the same meaning as $X\to -X$ and $Z\to -Z$ in Eq. (\ref{me}). It follows that  $Z' \to -Z'$ due to which  $\Theta\to \pi-\Theta$.
Therefore  the $k \to -k$ transformation is an isomorphism in the phase space.

Further symmetry can be found in the equations (\ref{me}) studying the classical map. 
The classical map contains the parameter $p$.  So we analyze the dependence on it which leads  to different simpler  equations.  first consider the system at $p=\frac{\pi}{2}$.
  It was studied by wide range of articles \cite{bs2,ms4,bl1,mgtg,nrfck,hks} in the literature. Due to newer symmetries, the shape of the mapping is simplified as follows
\begin{eqnarray}
\begin{array}{rl}
X' &=Z\cos(kX)+Y\sin(kX)\\
Y'&=Y\cos (kX)-Z\sin(kX)\\
Z'&=-X
\end{array}
\end{eqnarray}
 At small values $k$, the phase space is mainly covered by regular trajectories  (Fig. (\ref{figw-2})) at $k=0.07$.  The trivial fixed points is situated  at $(\Phi,\Theta)=(\pi/2,\pm\pi/2)$.
Increasing the value of $k$, the chaotic regions expands more and more in the phase space.
 For growing parameter value $k$ the phase space contains mainly chaotic sea with a few regular regions.

The map is studied for the value of the parameter $p=3\pi/2$.  It  is derived from $p=\pi/2$ by the the transformation $X'\to -X'$ and $Z'\to -Z'$. This means reflections about  $\Phi=0$ and $\Theta=\pi/2$ because  $\Phi \to -\Phi$ and $\Theta \to \pi-\Theta$. In the case of phase space, we also find such a behavior when we use these reflection.

The next $p$ value is chosen to be $\pi$, then the Equation (\ref{me})  of the classical map forms: 
\begin{eqnarray}
\begin{array}{rl}
X'&=Y\sin (kZ)-X\cos (kZ)\\
Y'&=Y\cos (kZ)-X\sin(kZ)\\
Z'&=-Z
\end{array}
\end{eqnarray}
In this case the fully developed chaos does not appear. The angle $\Theta$  is changing between $\cos^{-1}Z$ and $\pi-\cos^{-1}Z$ at a given initial value of $Z$. These  quantities are reflected for  $\pi/2$.

Th last instance is  $p=2\pi$ which we investigate
\begin{eqnarray}
\begin{array}{rl}
X'&=X\cos (kZ)-Y\sin(kZ)\\
Y'&=X\sin(kZ)+Y\cos(kZ)\\
Z'&=Z
\end{array}
\end{eqnarray}
In this situation  the fully developed chaos does not evolve for a given initial value $Z$ and the angle $\Theta$ equals to constant at $\cos^{-1}Z$.

\section{Quantum kicked rotor}\label{sec-rot}

The kicked rotor(QKR) plays an important role in the research of chaos. The Hamiltonian of this driven system is
\begin{eqnarray}\label{hrot}
H_R=\frac{1}{2I} P^2 + k\cos \Phi \sum_{n=-\infty}^{\infty} \delta (t-nT),
\end{eqnarray}
where $\Phi$ is the angle operator and $P$ is the angular momentum, canonically conjugate to $\Phi$ and $T$ is a periodic time.
The strength of the kick is denoted by $k$ and  $I$ is the moment of inertia and  the rotor operators satisfy the communication relation:
\begin{eqnarray}
[P,\Phi]=-i.
\end{eqnarray}
From the discrete dynamics, we get the angular operator and  angular momentum from driven to driven in the Heisenberg picture by these equations:
\begin{eqnarray}
\begin{array}{rl}
P'&=U_R^{\dagger}P U_R\\
\Phi'&=U_R^{\dagger}\Phi U_R,
\end{array}
\end{eqnarray}
where the Floquet uniter operator $U_R$ is defined by this term
\begin{eqnarray}
U_R= \exp\left(-i\frac{P^2}{2I}\right) \exp(-ik\cos \Phi)
\end{eqnarray}
The stroboscopic equations is as follows
\begin{eqnarray}\label{mer}
\begin{array}{rl}
P'&=P+k\sin\Phi \\
\Phi'& = \Phi+ P'/I.
\end{array}
\end{eqnarray}
The classical equation of motion following from Eq. (\ref{hrot})  in the literature it is known  as Chirikov's standard map \cite{ch1,grr,bszf} ($I=T=1$).

\begin{figure}
\begin{center}
\includegraphics[width=6.0cm]{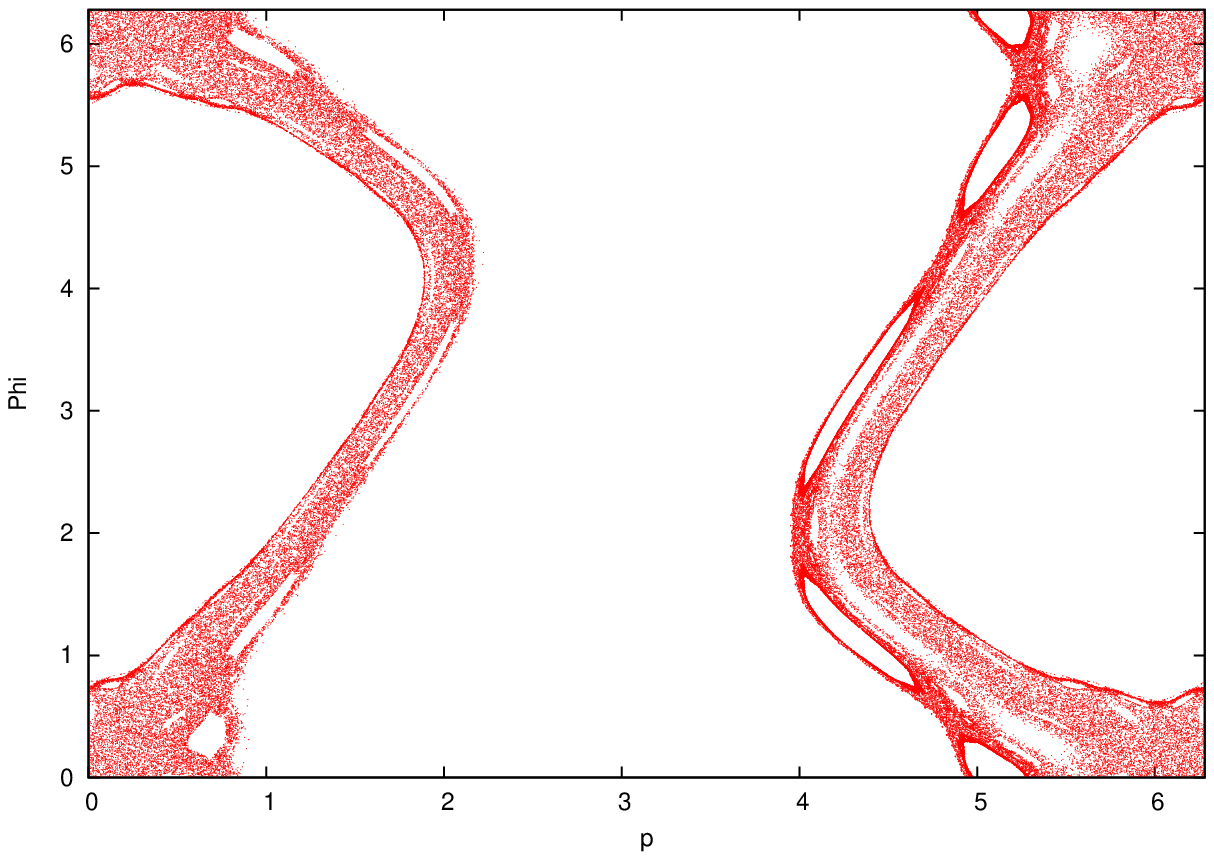}
\includegraphics[width=6.0cm]{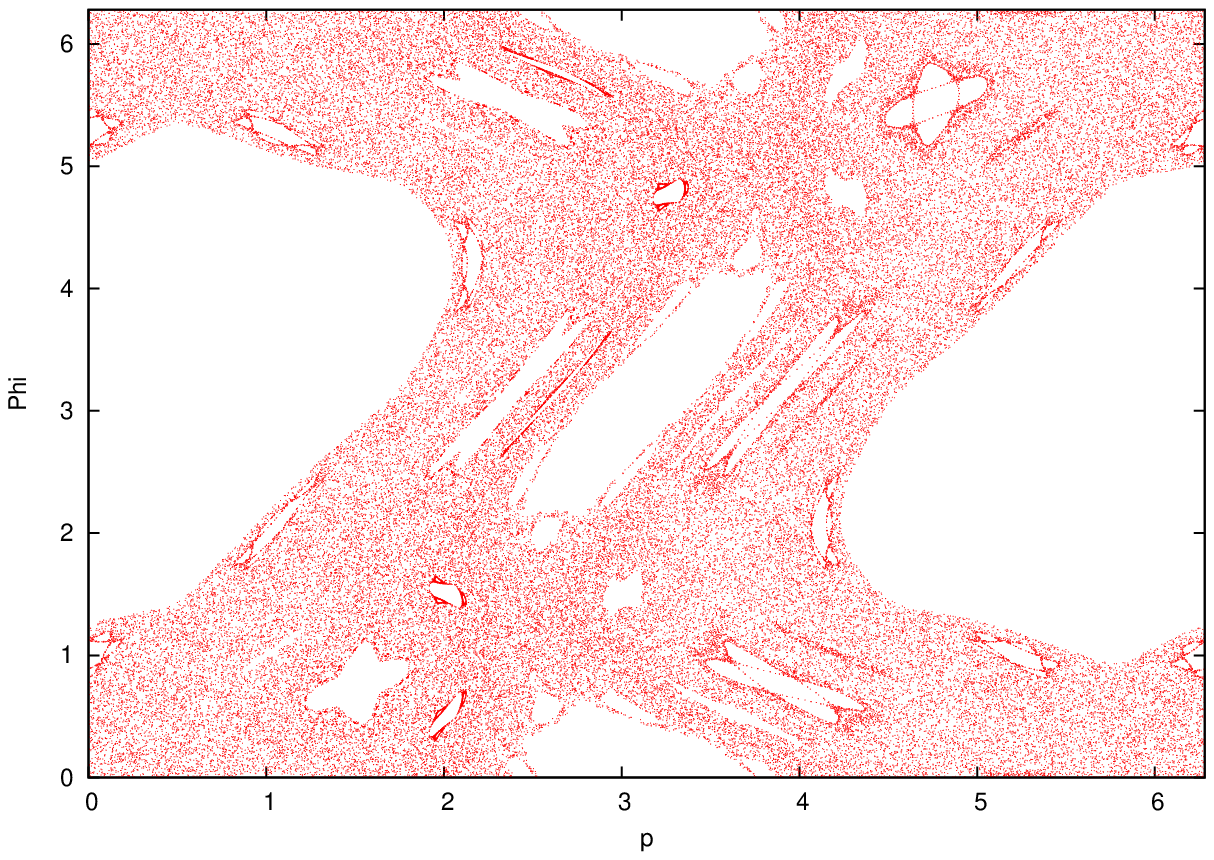}
\caption{Kicked rotor model $k=0.9$ $k=1.26$.}\label{fig-8}
\end{center}
\end{figure}

The surface of the rotor phase space is a  cylinder, $-\infty <P<\infty$, $0\le \Phi <2\pi$.
 It is seem from the stroboscopic equation that the model is invariant under $2\pi I$ translations in $P$ and $2\pi$ in $\Phi$, even though the model is not bounded in $P$.
 
Comparing the topology of the systems rotor and kicked top model this is dissimilar because  the kicked top model holds spherical phase space.   The classical rotor model is plotted in Fig. (\ref{fig-8}) for   $k=0.9$, $I=1$.

The classical phase space of the rotor model  contains  elliptic and hyperbolic fixed points and KAM tori (as circles) corresponding to  KAM theory and Poincar\'e-Birkhoff theorem.
   The KAM tori are invariant sets therefore the chaotic trajectories can not pass through  to evolve  hyperbolic fixed points. In the case of $k=0$ the model is a free rotor corresponding to  regular motion.  
  As the value of $k$ becomes larger, the KAM tori decompose into  cantori \cite{mms1,bk1}(invariant Cantor sets) and these are partly passable. There exists a critical nonlinearity $k_c$ with universal scaling properties \cite{g1,sk1}, where the final KAM tori  split the phase space into partition along the $P$ axis and the system becomes globally chaotic. 
    The classical KAM theory is extended to the  quantum sytems  \cite{grr}.
    
\paragraph{Classical rotor-limit}

\begin{figure}\label{fig-sph}
\begin{center}
\includegraphics[width=3.0cm]{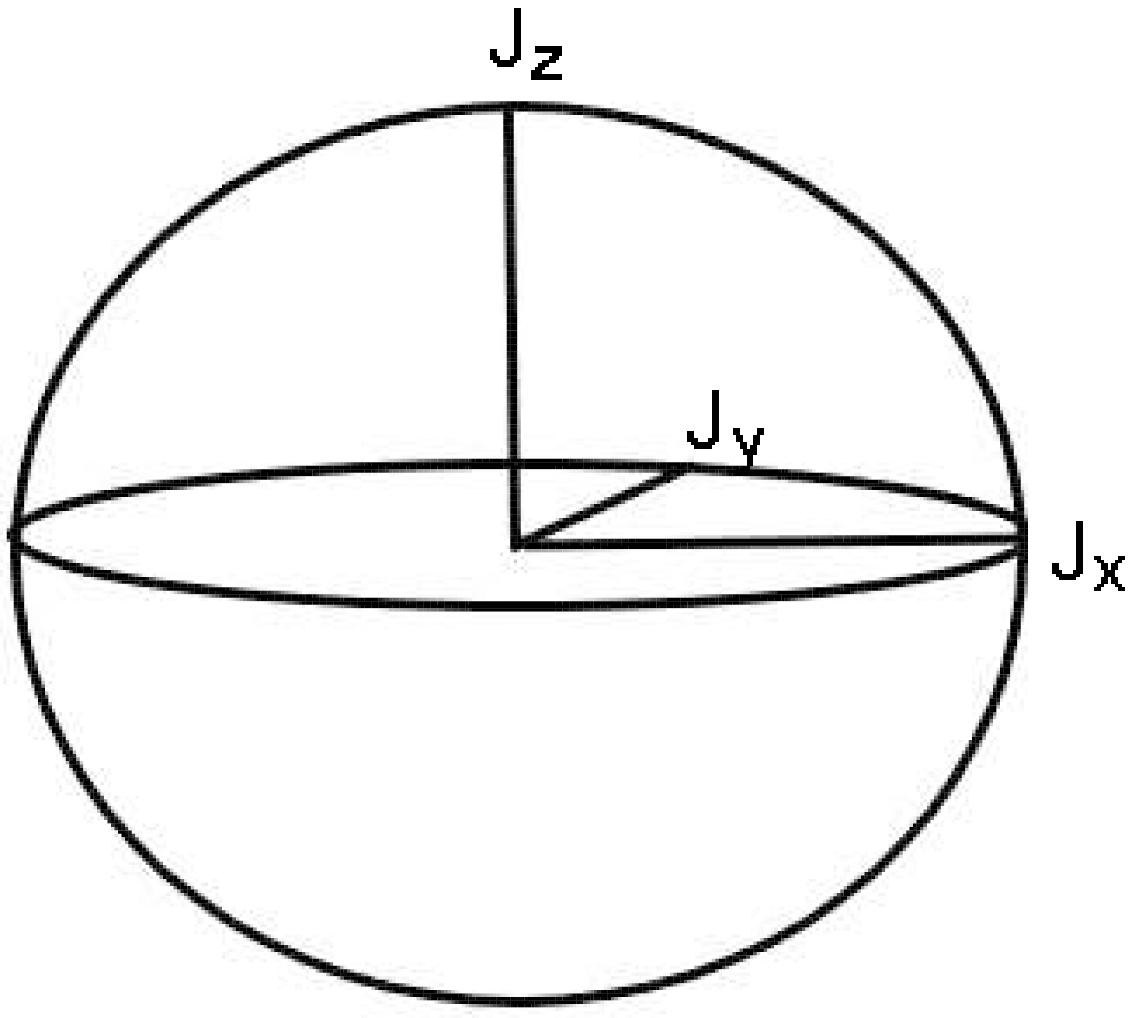}
\includegraphics[width=7.0cm]{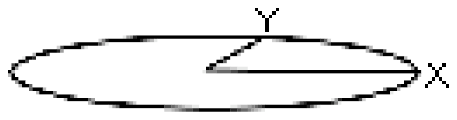}
\caption{(left)The magnitude of the angular momentum $J$ of the kicked top model is connserved quantity therefore it is displayed on a sphere. (right) The rotor limit is driven with rescaling $\alpha= k/j$, $\beta=j/I$, as $j\to \infty$. If we start in the equatorial waistband, the rescaling constrains the angular momentum to $X=\cos\Phi$, $Y=\sin \Phi$, $P/j$.}\label{fig-2}
\end{center}
\end{figure}

The time evolution of the rotor map can be risen from the kicked top model, if we bound the top to an equatorial waistband as plotted in Fig. (\ref{fig-2}). Then the precession frequency is decreased around the $x$-axis fulfilling the rescaling.
\begin{eqnarray}\label{ab}
\alpha=k/j, \;\;\;\beta=j/I,
\end{eqnarray}
where $j\to \infty$.
This means   the rotor-limit of the top kicked model.

We  begin in the equatorial waistband, this rescaling recricts the angular momentum to Fig.(\ref{fig-2})
\begin{eqnarray}\label{xyz}
X=\cos \Phi, \;\;\; Y=\sin \Phi \;\;\; Z=P/j.
\end{eqnarray}
If we replace the Eq (\ref{ab}) and (\ref{xyz}) in the kicked top map of Eq (\ref{me}), we get the kicked rotor map of Eq. (\ref{mer})\cite{hp1}.

\paragraph{Quantum rotor-limit}

Here we introduce the next rescaled operators:
\begin{eqnarray}\label{opk}
\hat{X}\equiv\hat{J}_x/j,\;\; \hat{Y}\equiv\hat{J}_y, \;\; \hat{P}\equiv\hat{J}_x
\end{eqnarray}
These operators fulfill the  communication relations of Eq. (\ref{komr}). Consider  $j\to \infty$ so that we may cat the $1/j^2$ terms:
\begin{eqnarray}\label{qkomr}
[\hat{X},\hat{Y}]=0,\;\; [\hat{Y},\hat{P}]=i\hat{X}, \;\; [\hat{P},\hat{X}]=i\hat{Y}.
\end{eqnarray}
These expressions fulfill the communication relations:
\begin{eqnarray}
\hat{X}=\cos\phi, \;\; \hat{Y}=\sin \Phi,\;\; \hat{P}=-i\frac{\partial}{\partial\Phi}.
\end{eqnarray}
These Equations (\ref{ab}) (\ref{opk}) and (\ref{qkomr}) are put in the top Hamiltonian of Eq. (\ref{ht}) than we obtain the rotor Hamiltonian of Eq. (\ref{hrot}).

\section{\bf Numerical approximation}\label{sec-num}

In this section we represent the statistical complexity of the kicked top and kicked rot model. 
This is a well signature of the chaotic features of these systems to show the mixed inner structure  and the time dependent evolution in the top model associating the behavior of qbits.

\paragraph{Statistical complexity}

The statistical complexity is introduced on the probability distribution yielding a statistical estimation  of the points in the phase space (section \ref{sec-stat}). 

The dynamical behavior of the systems is discussed in the next form \cite{af2}. 
The notation of measured sequence  is denoted by $y_1\dots y_n$ time series, where $y_i$ corresponds to measurement of the quantity $y$ at the time $t_i=t_0+iT$, $(T>0\in \mathbb{R})$.
The trajectory of length $n\in \mathbb{R}^d$ i.e. time series of the measurement is written by $\underline{x}^{(n)}$. The point of the orbit of the length $n$  is denoted by $x_k^{(n)}, (k=1\dots n)$ and the set $K$ contains  the points of some trajectories $x_k^{(n)}$ $(k=1\dots n)$. Let us consider a time  sequent of length $N'>>n$ .
 A given  series $x_k^{(n)}, (k=1\dots n)$ appears with probability $P(\underline{x}^{(n)})$ along the the sequences of length $N'$, where the corresponding set of discrete probability distribution $P\equiv\{p_1\dots p_{N'}\} $ , $p_i=P(\underline{x}^{n}_i)$ $(\sum_{i=1}^{N'} p_i=1)$, and $p_i>0\;\; \forall i$.

The driven systems can be derived numerically in different methods. 
We may determine a single very long trajectory of the periodically excited system or compute an ensemble of orbits. If the kicked model is periodic, the single long trajectory can be simulated stroboscopically in the three dimensional phase space. We select the smooth initial values in the past from the domain of the map  for the variables $x$, $y$ and $z$  at time $t=0$. The periodicity is $T=1$ and the  length of the trajectory is chosen as $N=10^4$. On this basis the value of the entropy, disequilibrium and statistical complexity are able to determine unambiguously.

 The motion of the (a)kicked top  and (b)kicked rotor map becomes on a surface  in the three dimensional phase space. The trajectories of the kicked top map are found on the sphere (section \ref{sec-top}) and the orbits of rotor map are located on cylinder (section \ref{sec-rot}). 
In the periodically driven model these systems depend on the parameter $k$, this means the strength of the excitation.

\begin{figure}
\begin{center}
\includegraphics[width=6.0cm]{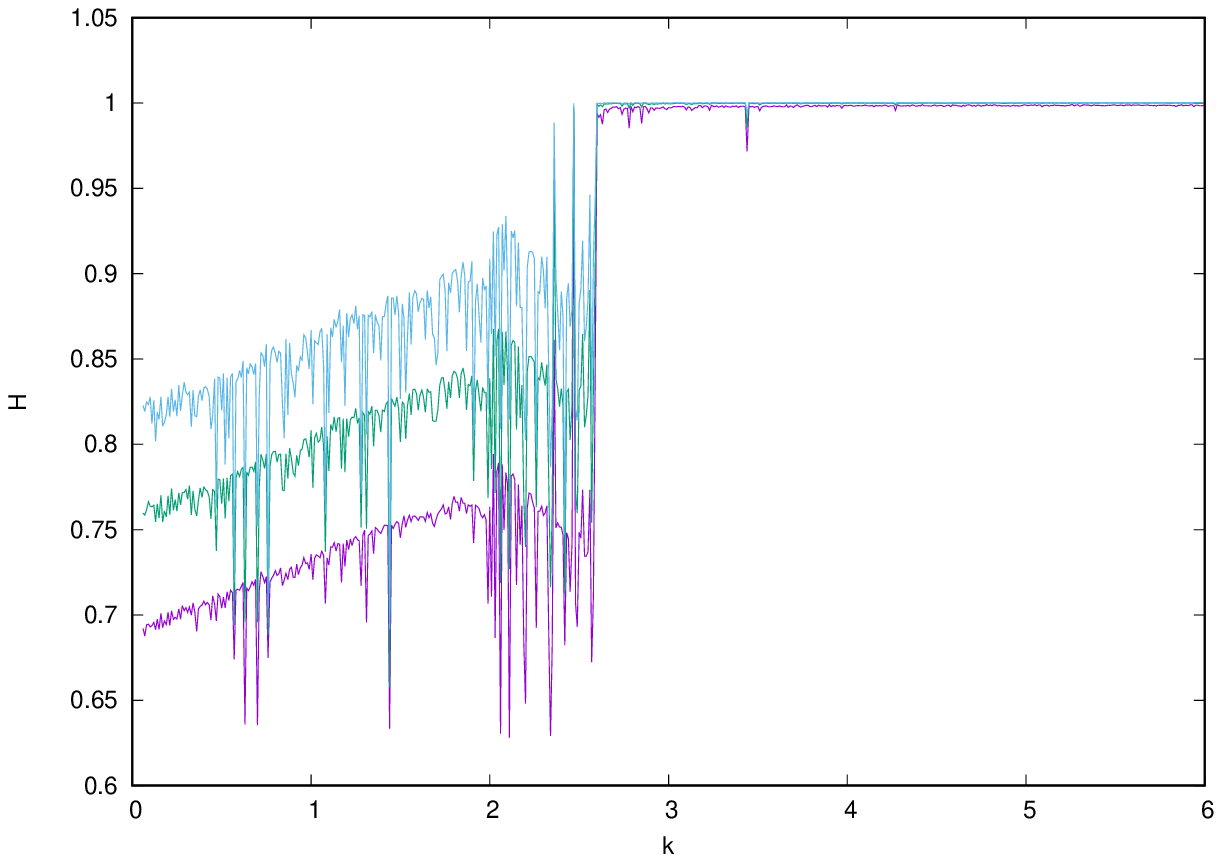}
\includegraphics[width=6.0cm]{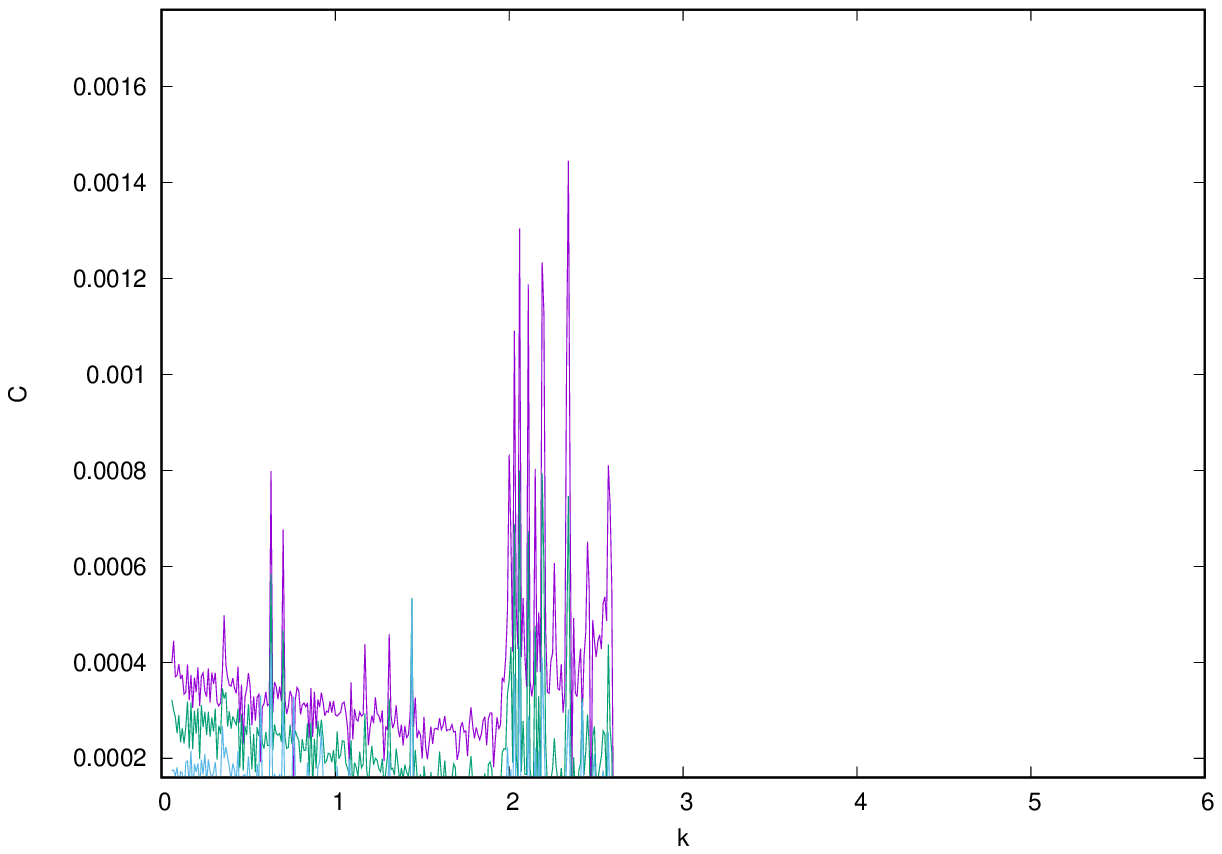}
\caption{Kicked top model (left):Entropy $H$ as a function strength of the driven $k$. (right): Statistical complexity $C$ depends on the driven parameter $k$.}\label{fig-3}
\end{center}
\end{figure}

\begin{figure}
\begin{center}
\includegraphics[width=6.0cm]{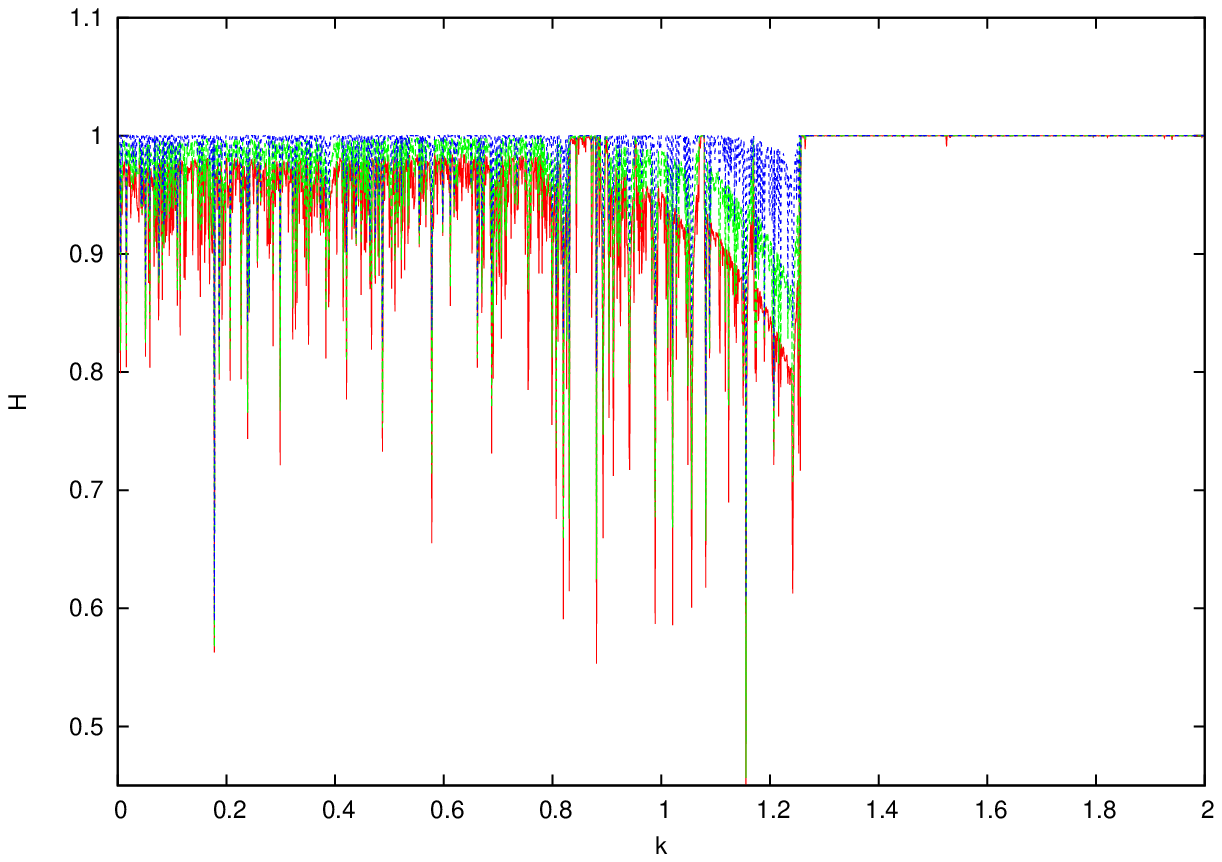}
\includegraphics[width=6.0cm]{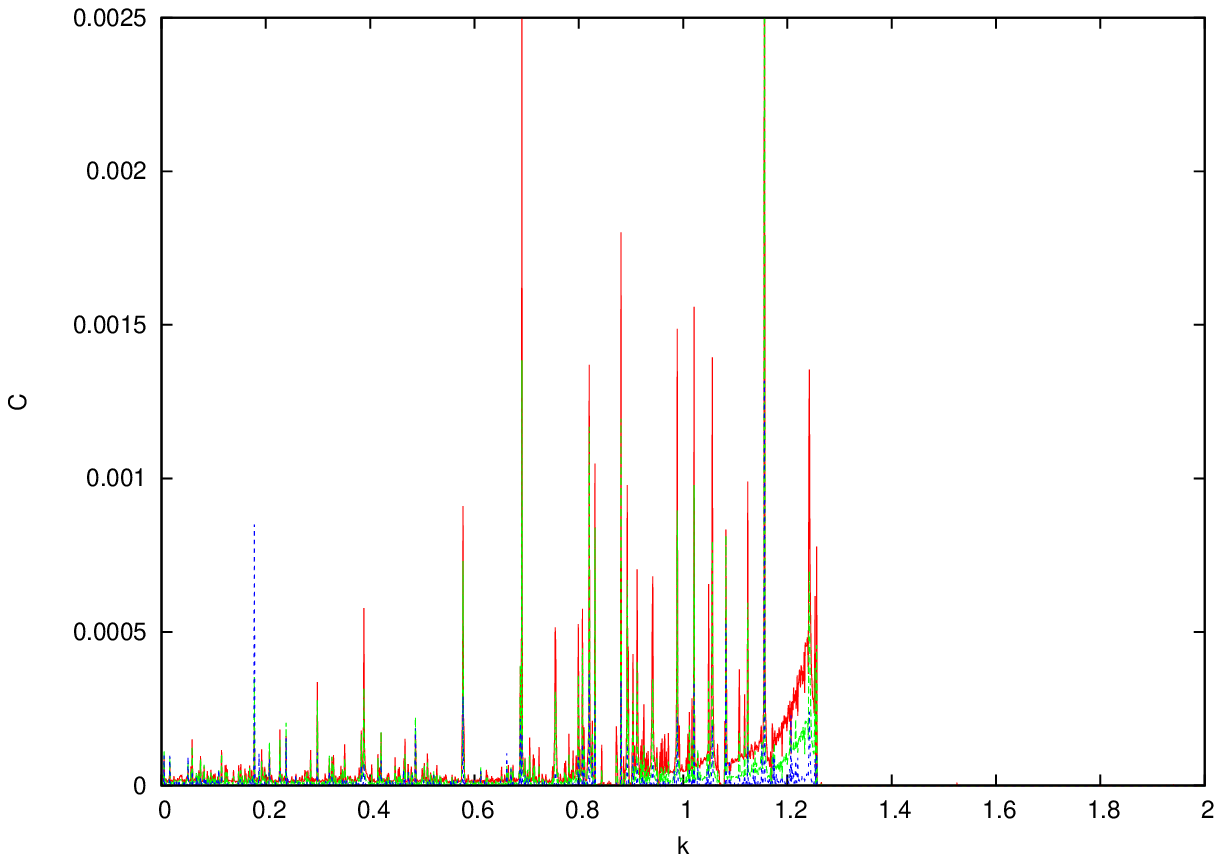}
\caption{kicked rotor model: (left): Entropy $H$ as a function strength of the driven $k$. (right): Statistical complexity $C$ depends on the driven parameter $k$.}\label{fig-4}
\end{center}
\end{figure}

 The chaotic behavior  turns up at the critical parameter value $k_c$ and the statistical complexity $C$  becomes to zero. The quantity of the kicked top map is $k_c=1.26$ and the value of the rotor map is $k_c=2.64$ ((a)Fig.\ref{fig-3}, (b)Fig\ref{fig-4}). 
Because the distance between the probability distribution $P$ and the uniform distribution $P_e$ tends to zero and the entropy approaches  1  at the equal probability points in the phase space.
  
\begin{figure}
\begin{center}
\includegraphics[width=6.0cm]{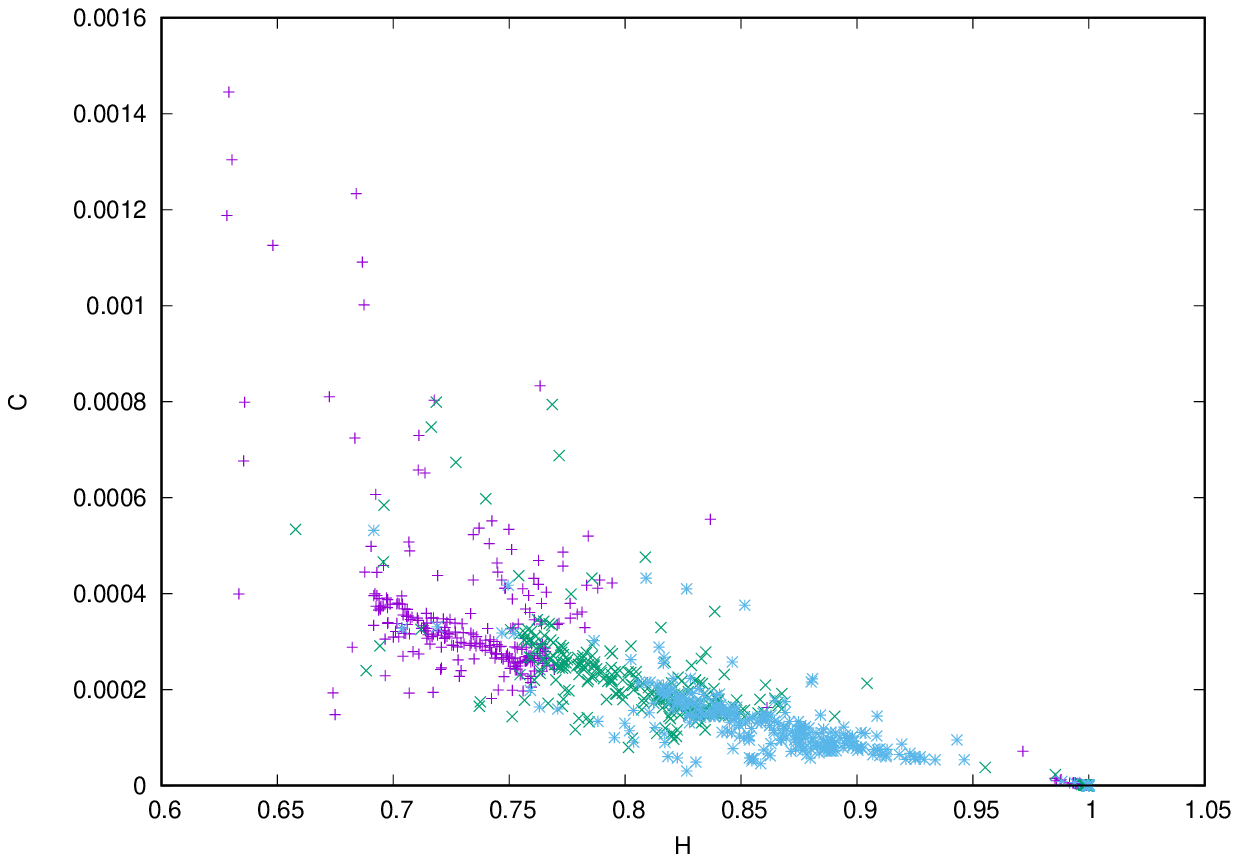}
\includegraphics[width=6.0cm]{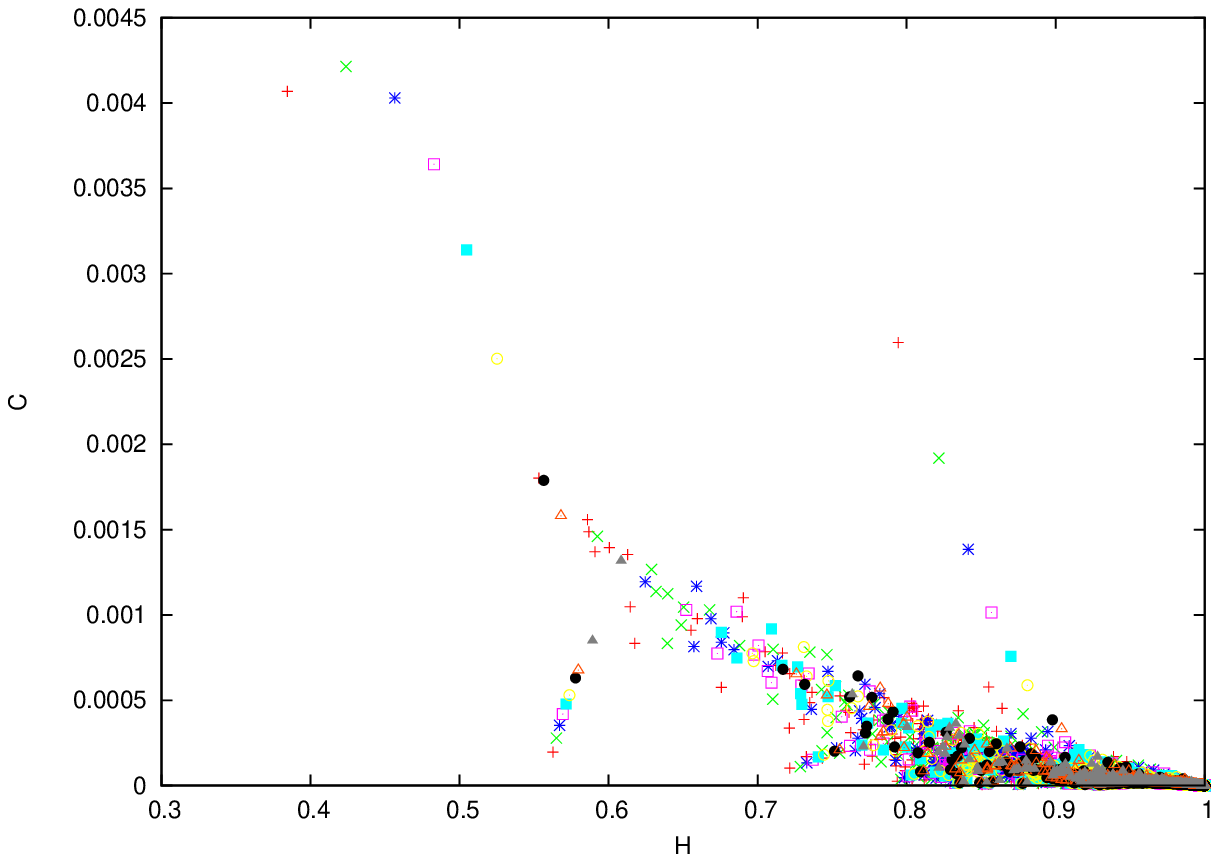}
\caption{The statistical complexity $C$ depends on the entropy $H$. left:kicked top model, right: kicked rotor system.}\label{fig-5}
\end{center}
\end{figure}

The parameter $k$ is extended to the neighbor of critical values (a)$k\in [0,6]$ (b)$k\in [0,2]$. 
Due to the increasing strong perturbation of the periodic driven systems, the statistical complexity $C$ decreases to zero at the same time the value of the entropy $H$ tends to 1  depending  on the parameter $k$, therefore the values $C$ and $H$ changing between extreme states $C_{max}$ at $H \sim 0$ and $C_{min}$ at $H\sim 1$ with the transition intervals. The chaotic behavior corresponds to the range, where the values is $C=0$ and $H=1$ (Fig.(\ref{fig-5})). For the calculation owing to the finite size of the simulation, the count accuracy becomes larger when the number of element $N$ increases. At the small value $k$ the periodically driven forcing does not have effect on the model i.e. the the motion of system  is regular on the surface. 

 The spectrum of the statistical complexity is finite and limiting but not inevitably a unique function of $H$ and there exists a range of values between a minimal value $C_{min}$ and a maximal value $C_{max}$ containing the inner structure (Fig.(\ref{fig-5})).

Because the number of points on the phase space is finite, $C$ as a function $H$  
shows scaling behavior, i.e. the bigger complexity associates  with less entropy with a larger discrete probability distribution. 
Since the probability distribution of element in the phase space is discontinuous  in the three-dimensional space, some complexity and disequilibrium values do not appear for certain entropy quantities.

\section{\bf Conclusion}

The model describing the qbits are  kicked top and rotor model, which are studied by statistical complexity in a finite probability distribution in a three-dimensional space considering the measure of the entropy and disequilibrium using the scaling behavior of these quantities. At the range of chaoticity the statistical complexity approximates zero and entropy goes to one.

We extended the parameter values of $k$ to the neighbor of the critical quantities studying the spectrum of the statistical complexity and the disequilibrium depending in the entropy. In the range of parameter $k$ we reached the dependence of the quantities $C$ and $H$, which  is acts as a periodical driven force.


\begin{thebibliography}{29}
\setlength{\parskip}{-3pt} 

\bibitem{ac}{ C. Adami, N.T. Cerf,}
Physical complexity of symbolic sequences,
{\em Physica D} {\bf 137} (2000) 62--69.

\bibitem{ap}{ C. Anteneodo, A.R. Plastino}, Some features of the L\'opez-Ruiz-Manchini-Calbet (LMC) statistical measure of complexity, {\em Physics Letters A} {\bf 223} (1996) 348--354.



\bibitem{bl1}{J. N. Bandyopadhyay and A. Lakshminarayan},
Entanglement production in coupled chaotic systems: Case of the kicked tops
 {\em Phys. Rev. E} {\bf 69} (2004) 016201.
 
\bibitem{bl3}J. N. Bandyopadhyay and A. Lakshminarayan,
Testing Statistical Bounds on Entanglement Using Quantum Chaos
 {\em Phys. Rev. Lett.}{\bf 89} (2002) 060402.
  
\bibitem{bas}{M. Basseville}, Information: Entropies, Divergences et Mayennes, (IRISA) Publication Interne {\bf 1020} (1996) (Campus Universitaire de Beaulieu, 35042 Rennes Cedex, France). 

\bibitem{bszf}{J. Bene, P. Szepfalusy, A. Fulop}
  Generic Dynamical Phase-transition IN Chaotic Hamiltonian-systems
  {\em Phys. Rev. A} {\bf 40} (1989)6719--6722.

  \bibitem{bk1}{D. Bensimon, L.P. Kadanoff},
  Extended chaos and disappearance of KAM trajectories
   {\em Physica} {\bf 13D} (1984)82.
  
 \bibitem{bp1}{P.M. Binder, N. Perry,}
  Comment II on: Simple measure of complexity.
   {\em Phys. Rev. E} {\bf 62} (2000)2998--2999.
   
    \bibitem{bs2}U. T. Bhosale and M. S. Santhanam,
 Signatures of bifurcation on quantum correlations: Case of the quantum kicked top
  {\em Phys. Rev. E} {\bf 95} (2017) 012216.
 
\bibitem{bs}{U. T. Bhosale and M. S. Santhanam}
Periodicity of quantum correlations in the quantum kicked top
{\em Phys. Rev. E} {\bf 98} (2018) 052228. 

\bibitem{av}{ G. Boffetta, M. Cencini, M. Falcioni, A. Vulpiani}, Predictability: a way to characterize complexity, {\em Phys. Reports} {\bf 356}(2002) 367--474.

\bibitem{bh1}{J. Briet, P. Harremoes,}
 Properties of classical and quantum Jensen-Shannon divergence. 
  {\em Phys. Rev. A} {\bf 79} (2009) 052311.
  
  
\bibitem{cl}{X. Calbet, R. L\'opez-Ruiz}, Tendency towards maximum complexity in a nonequlibrium isolated system, {\em Phys. Rev. E} {\bf 63} 066116.
   
  \bibitem{csagj} S. Chaudhury, A. Smith, B. E. Anderson, S. Ghose, and P. S.
Jessen,
Quantum signatures of chaos in a kicked top
 {\em Nature} {\bf 461} (2009) 768.
  
  \bibitem{ch1}{B.V. Chirikov}
  A universal instability of many-dimensional oscillator systems
  {\em Phys. Rep.} {\bf 52} (1979) 265.
  
\bibitem{cfs1}{J.P. Crutchfield,D.P. Feldman,C.R. Shalizi}
  Comment I on: simple measure of complexity. 
  {\em Phys.Rev. E} {\bf 62} (2000) 2996--2997. 

\bibitem{cy} {J. P. Crutchfield, K. Young,}
Inferring statistical complexity, {\em Phys. Rev. Lett.} {\bf 63} (1989) 105.

\bibitem{fc1}{ D.P. Feldman, J.P. Crutchfield,}
 Measures of statistical complexity: Why?
  {\em Phys. Lett. A} {\bf 238} (1998)244--252.
  
\bibitem{fpp}{ G.L. Ferri, F. Pennini, A. Plastino,}  LMC-complexity and various chaotic regime,
 {\em Physics Letters A} {\bf 373} (2009) 2210--2214.
 
\bibitem{fmt1}H. Fujisaki, T. Miyadera, and A. Tanaka,
Dynamical aspects of quantum entanglement for weakly coupled kicked tops
{\em Phys. Rev. E} {\bf 67}, (2003)066201.
  
\bibitem{af2} { \'A. F\"ul\"op,}
Estimation of the Kolmogorov entropy in the generalized number system, {\em Annales Univ. Sci. Budap est Sect. Comp.} {\bf 40} (2013) 245--256.

\bibitem{afs}{ \'A. F\"ul\"op,}
Statistical complexity and generalized number system, Acta Univ. Sapientiae, Informatica {\bf 6} (2) (2014) 230--251.

\bibitem{grr}{T. Geisel, G. Radons, and J. Rubner}
Kolmogorov-Arnol'd-Moser Barriers in the Quantum Dynamics of Chaotic Systems
{\em Phys Rew. Letters} {\bf 57} (1986) 2883.

\bibitem{gsjls}S. Ghose, R. Stock, P. Jessen, R. Lal, and A. Silberfarb,
Chaos, entanglement, and decoherence in the quantum kicked top
{\em Phys. Rev. A }{\bf 78} (2008) 042318.


\bibitem{cghl}{ C.M. Gonzalez, H.A Larrondo, O.A. Rosso,}Statistical complexity measure of pseudorandom bit generators,
{\em Physica A} {\bf 354} (2005) 281.

\bibitem{gpijt}{ P. Grassberger,} Toward a Quantitative Theory of Self-Generated Complexity, {\em Int. Journ. Theor. Phys.} {\bf 25} (1988) 907--938.

   \bibitem{g1}{J.M. Greene} 
  A method for determining a stochastic transition
  {\em J. Math. Phys.} {\bf 20} (1979) 1183.

\bibitem{hks}{F. Haake, M. Kus, and R. Scharf}, 
Classical and quantum chaos for a kicked top
 {\em Z. Phys. B} {\bf 65} (1987) 381.

\bibitem{hp1}{F. Haake and D. L. Shepelyansky,}
 The kicked rotator as a limit of the kicked top,
 {\em EPL (Europhys. Lett.)} {\bf 5} (1988) 671.

\bibitem{ank} { A.N. Kolmogorov}, Entropy per unit time as a metric invariant of automorphism, {\em Doklady of Russian Academy of Sciences}, {\bf 124} (1959) 754--755.

\bibitem{entr}{ A.M. Kowalski, M.T. Martin, A. Plastino, O-A. Rosso, M. Casas}, Distances in Probability space and the statistical complexity setup, {\em Entropy} {\bf 13} (2011) 1055--1075.

\bibitem{kl1} {S. Kullback, R.A Leibler,} 
On Information and Sufficiency
{\em Ann. Math. Stat.} {\bf 22} (1951)79–86.

 \bibitem{ms4}{M. Kumari and S. Ghose}
 Quantum-classical correspondence in the vicinity of periodic orbits
{\em Phys. Rev. E} {\bf 97} (2018) 052209.

\bibitem{l3}A. Lakshminarayan,
Entangling power of quantized chaotic systems
 {\em Phys. Rev. E} {\bf 64} {2001} 036207.

\bibitem{lmpr}{P.W. Lamberti, M.T. Martin, A. Plastino, O.A. Rosso,} 
Intensive entropic nontriviality measure.
{\em Physica A} {\bf 334} (2004) 119–131.

\bibitem{lz}{A. Lempel, J. Ziv} 
On the complexity of finite sequences, IEEE Trans. Inform Theory {\bf 22} (1976) 75--81.

\bibitem{lm2}M. Lombardi and A. Matzkin,
Entanglement and chaos in the kicked top
 {\em Phys. Rev. E} {\bf 83}, {2001} 016207 (2011).

\bibitem{lmc}{ R. L\'opez-Ruiz, H.L. Mancini, X. Calbet,}  A statistical measure of complexity,
 {\em Phys. Letters A} {\bf 209} (1995) 321--326.
  
\bibitem{lo} { M. Lovallo, V. Lapenna, L. Telesca,} Transitionmatrix analysis of earthquake magnitude sequences {\em Chaos,Soliton and Fractals} {\bf 24} (1) (2005) 33--43.
 
  \bibitem{mms1}{R.S. Mackay, J.D. Meiss, I.C. Shepelyanski}
  Transport in Hamiltonian systems
  {\em Physica} {\bf 13D} (1984) 55.
  
\bibitem{mgtg}{V. Madhok, V. Gupta, D.A. Trottier, and S. Ghose,}
Signatures of chaos in the dynamics of quantum discord
 {\em Phys. Rev. E} {\bf 91} (2015) 032906.

\bibitem{mdl1}V. Madhok, S. Dogra, and A. Lakshminarayan,
Quantum correlations as probes of chaos and ergodicity
{\em Opt. Commun.}{\bf 420}(2018) 189.
 
\bibitem{mpr}{ M.T. Martin, A. Plastino, O.A. Rosso,}  Statistical complexity and disequilibrium, {\em Physics Letters A}  
{\bf 311} (2003) 126--132.

\bibitem{mpo}{ M-T. Martin, A. Plastino, O.A. Rosso,}  Generalized statistical complexity measures: Geometrical and analytical properties, {\em Physica A } {\bf 369} (2006) 439--462.

\bibitem{ms3}P. A. Miller and S. Sarkar,
Signatures of chaos in the entanglement of two coupled quantum kicked tops
 {\em Phys. Rev. E} {\bf 60} (1999) 1542.

\bibitem{mc}{H. Ming-Liang and X. Xiao-Qiang, Chin.}
Mixedness of the N-qubit states with exchange symmetry
 {\em Phys. B} {\bf 17} (2008) 3559.

\bibitem{nrfck}{C. Neill, P. Roushan, M. Fang, Y. Chen, M. Kolodrubetz, Z.
Chen, A. Megrant, R. Barends, B. Campbell, B. Chiaro et al.,}
Ergodic dynamics and thermalization in an isolated quantum system
{\em Nat. Phys.}  {\bf 12} (2016) 1037.


\bibitem{plq}{A. Piga, M. Lewenstein, and J. Q. Quach}
Quantum chaos and entanglement in ergodic and nonergodic systems
{\em Phys. Rev. E} {\bf 99} (2019) 032213.

\bibitem{pp1} {A.R. Plastino, A. Plastino,}
Symmetries of the Fokker-Plank equation and Fisher-Frieden arrow of time. 
{\em Phys. Rev. E} {\bf 54} (1996)4423--4326.

\bibitem{rlmpf}  {O.A. Rosso, H.A. Larrondo, M.T. Martin, A. Plastino, M.A. Fuentes,}
 Distinguishing noise from chaos. 
 {\em Phys. Rev. Lett.} {\bf 99} (2007) 154102.   

\bibitem{rmlmp} {O.A. Rosso, L. De Micco, H.A. Larrondo, M.T.  Martin, A. Plastino,}
 Generalized statistical complexity measure. 
 {\em Int. J. Bif. Chaos} {\bf 20}(2010)775–785.
  
\bibitem{rlp2}J. B. Ruebeck, J. Lin, and A. K. Pattanayak,
Entanglement and its relationship to classical dynamics
 {\em Phys. Rev. E} {\bf 95} (2017)062222.


\bibitem{cs} { C.E. Shannon},
  The Mathematical Theory of Communication,
 {\em Bell System Technical Journal}, {\bf 27} (1948) 379--423, 623--656.
    
  \bibitem{sk1}{S.J. Shenker, L.P. Kadanoff} 
  Critical behavior of a KAM surface: I. Empirical results
  {\em J. Stat. Phys.} {\bf 27} (1982) 631.

\bibitem{plx} { J.S. Shiner, M. Davison, P.T. Landsberg,} Simple measure for complexity, 
{\em Phys. Rev. E} {\bf 59}(2)(1999)1459--1464.

\bibitem{sdl1}{J.S. Shiner, M. Davison,P.T Landsberg,}
 Replay to comments on: simple measure for complexity.
{\em Phys. Rev. E} {\bf 62} (2000) 3000--3003.

\bibitem{sg3}G. Stamatiou and D. P. K. Ghikas,
Quantum entanglement dependence on bifurcations and scars in non-autonomous systems. The case of quantum kicked top
{\em Phys. Lett. A} {\bf 368} (2007) 206.

\bibitem{ct}{ C. Tsallis,}
Possible generalization of Boltzmann-Gibbs statistics,
{\em J. Stat. Phys.} {\bf 52} (1988) 479.

\bibitem{wwaks}{R. Wackerbauer, R.A. Witt, H. Atmanspacher, J. Kurths, H. Scheingraber,} A comparative classification of complexity-measures. 
{\em Chaos Solitons Fractals} {\bf 4} (1994) 133--173.

\bibitem{wgsb}{X. Wang, S. Ghose, B. C. Sanders, and B. Hu}
Entanglement as a signature of quantum chaos
{\em Phys. Rev. E} {\bf 70} (2004) 016217.

\bibitem{wkw}{ W.K. Wootters},
Statistical distance and Hilbert space,
{\em Phys. Rev. D} {\bf 23} (1981) 357.

\bibitem{zs}{R. Zarum and S. Sarkar}
Quantum-classical correspondence of entropy contours in the transition to chaos
{\em Phys. Rev. E} {\bf 57} (1998) 5467. 


\end{thebibliography}
\end{document}